\documentclass[10pt,a4paper]{article}

\usepackage{graphicx}
\usepackage{dcolumn}
\usepackage{bm}

\newcommand{\tr}[1]{\textrm{Tr}\left( #1 \right)}

\newcommand{\de}{\partial}
\newcommand{\sech}[1]{\textrm{sech}\left(  #1\right)}
\newcommand{\eq}[2]{\begin{equation} \label{#1} #2 \end{equation}}
\newcommand{\eps}{\epsilon}

\newcommand{\cc}{\textrm{c.c.}}

\newcommand{\sx}{\sigma_{x}}

\newcommand{\sz}{\sigma_{z}}

\newcommand{\sxplus}{\sigma_{x}^{+}}

\newcommand{\EE}{\mathbf{E}}
\newcommand{\BB}{\mathbf{B}}

\newcommand{\PP}{\mathbf{P}_{L}}

\title{\bf Dynamics of light propagation in spatiotemporal dielectric structures}

\author{Fabio Biancalana, Andreas Amann, Alexander V. Uskov, Eoin P. O'Reilly
\\
Tyndall National Institute, Lee Maltings, Cork, Ireland}

\begin{document}

\maketitle

\begin{abstract}
Propagation, transmission and reflection properties of linearly
polarized plane waves and arbitrarily short electromagnetic pulses
in one-dimensional dispersionless dielectric media possessing an
arbitrary space-time dependence of the refractive index are studied
by using a two-component, highly symmetric version of Maxwell's
equations. The use of any slow varying amplitude approximation is
avoided. Transfer matrices of sharp nonstationary interfaces are
calculated explicitly, together with the amplitudes of all secondary
waves produced in the scattering. Time-varying multilayer structures
and spatiotemporal lenses in various configurations are investigated
analytically and numerically in a unified approach. Several new
effects are reported, such as pulse compression, broadening and
spectral manipulation of pulses by a spatiotemporal lens, and the
closure of the forbidden frequency gaps with the subsequent opening
of wavenumber bandgaps in a generalized Bragg reflector.
\end{abstract}

\section{Introduction}

Propagation of electromagnetic waves through stationary
inhomogeneous media is a subject that has been extensively studied
in the past \cite{inhomogeneous,pochiyeh}.

The physics of moving dielectric media
\cite{jackson,landau,yeh,saca,huang} is attracting renewed
attention, mainly due to recent works by Visser \cite{visser} and
Leonhardt and Piwnicki \cite{leo}. Particularly important is the
formal equivalence between the equations of general relativity and
the equations of light propagation in arbitrarily moving dielectric
media, which has been fully recognized in the early literature
\cite{defelice}.

Much less attention, however, has been devoted to the case of light
propagation through an inhomogeneous medium which possesses a
time-dependent refractive index, and in particular light scattering
by nonstationary interfaces between two media with different
refractive indices. This must be distinguished from the above case
of a moving medium, in that the nonstationary interfaces may possess
arbitrary velocity $v$, in the range $0<|v|<\infty$. This well-known
fact \cite{ostrovsky} is not in conflict with special relativity
because the medium itself is immobile, and only the interfaces of
the transition regions of the refractive index change in time. This
apparent motion of the interfaces is due to the time-varying nature
of the refractive index, since it is not connected to the real
motion of any physical particle or wave, and is not restricted by
any limiting velocity, so that the function $n(z,t)$ that describes
the spatiotemporal variations of the refractive index is completely
arbitrary. In particular Lorentz transformations, which of course
play a crucial role in the physics of moving media, are not
applicable to the physics of media with a time-varying refractive
index. For example, the field amplitudes of the waves generated
inside the latter media are not found via Lorentz boosts, and the
constitutive relations between the dielectric displacement and the
electric field are fixed and not subject to Lorentz transformations,
in contrast with the physics of moving media, described, for
instance, in Refs. \cite{jackson,landau}.

The first attempt to develop a theory for the very special case of
plane wave propagation in a homogeneous medium with a sudden
step-like temporal variation of refractive index has been given by
Morgenthaler in 1958 \cite{morgenthaler}, but only after more than a
decade this work has been noticed and other aspects of propagation
in the same system have been explored by Felsen and Whitman in 1970
\cite{felsen} and by Fante in 1971 \cite{fante}. A more recent
review of the salient features of the optics of nonstationary media,
with particular emphasis on plasma physics applications, has been
given by Shvartsburg in 2005 \cite{nonstationary}.

The purpose of the present paper is to study the effect of an {\em
arbitrary} space- and time-dependence of refractive index $n(z,t)$
on the propagation of electromagnetic waves and pulses in media
without dispersion. We shall provide a unified theoretical approach,
with which we are able to study the scattering of light by
nonstationary interfaces moving at arbitrary velocities. Light
propagation through media with a small number of nonstationary
interfaces has been investigated by a variety of experimental
techniques
\cite{berezhiani,avitzour,wilks,lampe,hashimshony,bessarab}, as
discussed further in the final discussion and conclusions section.
Our results here include these cases but also address the behavior
that could be achieved by further extension to an arbitrary number
of well-controlled nonstationary interfaces. Our aim is to
demonstrate the potential of nonstationary dielectrics for light
control and manipulation.

The paper is organized as follows.

In Section \ref{sec:analyticalresults} we construct the basic
theoretical tools that are necessary for the analysis and the
understanding of the building blocks of spatiotemporal dielectric
structures. First of all, in subsection \ref{sec:main-equation} our
master equation [Eq.(\ref{main})] is derived from Maxwell's
equations, and its geometrical interpretation in terms of forward-
and backward-moving fields is given. After that, in subsection
\ref{matrices} the interface transfer matrix for the general case is
calculated explicitly, while in subsection \ref{sec:calc-prop-matr}
the expression for free propagation when the fields are far from
interfaces is derived, and a criterion for defining sharp interfaces
is found. In subsection \ref{sec:plane-wave-expansion} we use
special orthogonal variables to introduce plane wave expansion of
arbitrary fields, which allows us to find the transfer propagation
matrix in Fourier space.

In Section \ref{applications} we describe three fundamental
applications of the concept expressed in Section
\ref{sec:analyticalresults}. In subsection
\ref{nonstationaryinterface} the reflection and transmission
coefficients of several important configurations of nonstationary
interfaces are derived, and numerical simulations of pulse
scattering from interfaces are provided in order to confirm the
theoretical expressions. In subsection \ref{sec:photoniccrystals} we
discuss nonstationary multilayer structures, and provide the
analogue of quarter wavelength condition for a generalized
nonstationary Bragg reflector, and the transformation of the
bandstructure and the bandgaps under rotations of the $(z,ct)$-plane
are examined. In subsection \ref{spacetimelenses} we deal with a
special device, the spatiotemporal analogue of a lens. We show,
including numerical simulations, how such a lens can be used to
perform pulse compression or broadening, with the consequent
manipulation of pulse frequency and wavenumber.

The results presented in Sections \ref{sec:analyticalresults} and
\ref{applications} are obtained by solving the full Maxwell
equations. We derive in Section \ref{slow} an exact equation to
describe the evolution of the pulse envelope. We consider how this
equation is modified if we introduce a slowly varying envelope
approximation (SVEA) in time. Numerical simulations are then
presented to show that the SVEA introduces errors in the detail of
pulse propagation for the problems we are considering, due to the
neglect of effects associated with sharp spatiotemporal interfaces.
Finally we summarize our conclusions in Section \ref{conclusions}.

\section{Analytical Results}
\label{sec:analyticalresults}

\subsection{Main Equation}
\label{sec:main-equation}

Let us start by assuming that we have an electromagnetic wave, with
its electric and magnetic fields $\EE=[E(z,t),0,0]^{T}$ and
$\BB=[0,B(z,t),0]^{T}$ linearly polarized along the directions
$\hat{x}$ and $\hat{y}$ respectively. The $x$-component of the
electric field and the $y$-component of the magnetic field depend on
the variable $z$ only, because in conditions of normal incidence we
assume that any change of the time-dependent refractive index occurs
along the propagation direction $\hat{z}$. The linear polarization
$\PP$ of the medium will be written in the form
$\PP=\chi(z,t)\EE=[n(z,t)^{2}-1]\EE$, where $\chi$ is the linear
susceptibility of the non-magnetic medium and $n(z,t)$ is the linear
space-time dependent refractive index, which is assumed for
simplicity to be independent of the frequency $\omega$ and to be
real. This model for the susceptibility is obviously simplified, but
for the purposes of the present paper the frequency dispersion of
the refractive index is not essential and will be disregarded for
sake of clarity. Here and in the following we choose to use the very
convenient Heaviside-Lorentz units system \cite{jackson}, in which
$\EE$, $\BB$ and $\PP$ are measured in the same physical units
(V/m), and there are no extra multiplying factors $4\pi$, $\eps_{0}$
and $\mu_{0}$ in Maxwell's equations. Maxwell's equations for the
scalar quantities $E(z,t)$ and $B(z,t)$ can be written in the
following form ($c$ is the speed of light in vacuum):
\eq{maxwell}{\de_{t}(n^2E)+c\de_{z}B=0,\qquad \de_{t}B+c\de_{z}E=0,}
where the two divergence Maxwell's equations are automatically
satisfied due to the chosen polarization and the normal incidence
assumption.

It is possible to cast Eqs.(\ref{maxwell}) in a different form,
which turns out to be much more convenient for the analysis of
time-varying photonic crystals and lenses, using the fact that the
problem under consideration is (1+1)-dimensional. Defining {\em
local} right- and left-moving fields, $F^{\pm}\equiv(nE\pm B)/2$, we
can write Eqs.(\ref{maxwell}) in the following compact form:
\begin{equation}
  \label{main}
  \left[ \hat{\de}+\sx(\hat{\de}\ln{n})\sxplus  \right]\Psi=0,
\end{equation}
where $\Psi\equiv[F^{+},F^{-}]^{T}$ (with the superscript $T$ we
indicate vector transposition),
and \eq{pauli}{\sigma_{0}=\left(\begin{array}{cc} 1 & 0 \\
0 & 1
\end{array} \right),\qquad\sigma_{x}=\left(\begin{array}{cc} 0 & 1 \\ 1 &
0 \end{array} \right)\qquad\sigma_{z}=\left(\begin{array}{cc} 1 & 0
\\ 0 & -1
\end{array} \right),} are the identity matrix and
the first and third (real) Pauli matrices. Moreover,
$\sigma^{\pm}_{j}\equiv(\sigma_{0}\pm\sigma_{j})/2$ are the Pauli
projectors ($j=\{x,z\}$), and $\hat{\de}$ is the derivative operator
\eq{derivative}{\hat{\de}\equiv\frac{n}{c}\sigma_{0}\de_{t}+\sigma_{z}\de_{z}\equiv
\left(\begin{array}{cc} \de^{+} & 0 \\ 0 & \de^{-}
\end{array}\right),} with $\de^{\pm}\equiv (n/c)\de_{t}\pm\de_{z}$.
$(\de^{+},\de^{-})$ can be thought of as basis vectors for a tangent
space $\mathcal{TM}$ of the two-dimensional manifold $\mathcal{M}$
formed by the $(z,ct)$-plane equipped with a space-time dependent
refractive index function $n(z,t)$ \cite{nakahara}. The
spatiotemporal evolution of the dielectric structure is only
contained in the logarithmic term $\hat{\de}\ln{n}$, which couples
the forward- and the backward-propagating components of the
electromagnetic field. All conclusions of this paper are based on an
analysis of the main Eq.(\ref{main}).

\begin{figure}
\includegraphics{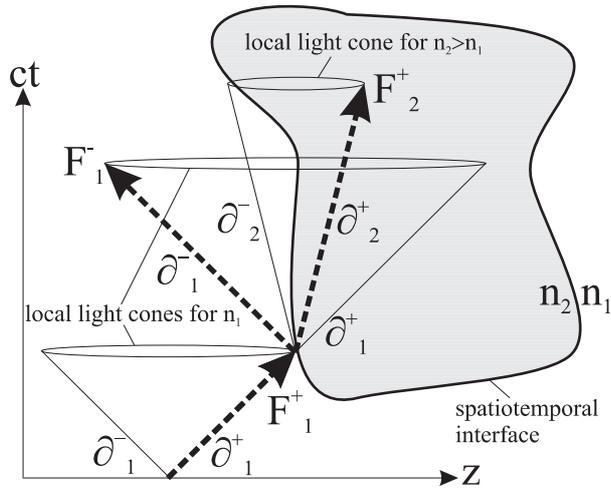}
\caption{\label{geometry} The geometry of propagation of fields
  $F^{\pm}_{j}$ along the directions determined by $\de^{\pm}_{j}$ is
  shown, where $j=\{ 1,2 \}$ refers to the refractive index
  $n_{j}$. The grey area represents the spatiotemporal region of
  refractive index change in the plane $(z,ct)$. For this simple
  example, the external and the internal regions have uniform indices
  $n_{1}$ and $n_{2}$ respectively. Bold dashed arrows indicate the
  propagating fields, generated by the scattering of $F^{+}_{1}$ from
  an interaction point on the interface. Local light cones are also
  indicated. Assuming that $n_{2}>n_{1}$, wider light cones are
  associated with index $n_{1}$, whereas narrower light cones are
  associated with index $n_{2}$. Only space-time variations of refractive index are able
  to generate coupling between forward and backward waves in Eq.(\ref{main}), so that
  in the case shown in the figure only the interface is able to scatter an incident electromagnetic wave.}
\end{figure}

One can appreciate the remarkably high level of symmetry between
space and time contained in Eq.(\ref{main}) with the following
arguments. The propagation of the local field $F^{+}$ occurs only
along the (local) direction specified by $\de^{+}$, while the field
$F^{-}$ evolves only along $\de^{-}$. Therefore  $\de^{+}$ and
$\de^{-}$ determine, at each point of the $(z,ct)$-plane, the local
light cone along which the forward and backward components of the
electric field propagate. The geometrical meaning of our
construction based on Eq.(\ref{main}) is depicted in
Fig.~\ref{geometry}, where examples of the local light cones defined
by the directions determined by $\de^{\pm}$ and the propagating
forward/backward fields $F^{\pm}$ are shown (see also caption for
further explanation of the geometrical meaning of our construction).

Contrary to a widespread approach \cite{scalora}, we do not make use
of any complex slowly varying variable, such as a field envelope,
and therefore Eq.(\ref{main}) remains valid for all regimes of
propagation, and all quantities appearing in Eq.(\ref{main}) are
real. A slow variable version of Eq.(\ref{main}) will be examined in
Section \ref{slow}, and a comparison between the exact approach of
Eq.(\ref{main}) and the approximate one of Eq.(\ref{scalorasveat})
will be presented.

Note that the coupling between $F^{+}$ and $F^{-}$ is only induced
via the logarithmic term $\hat{\de}\ln{n}$ in Eq.(\ref{main}), so
that only space-time variations of the refractive index are
responsible for coupling forward and backward waves during
propagation. We can therefore identify two basic cases in which the
analytical solution of Eq.(\ref{main}) is straightforward, namely
the case where $n$ changes abruptly from $n_1$ to $n_2$ along a line
in the $(z,ct)$-plane, and the case of a homogeneous and static
medium, where Eq.(\ref{main}) simply becomes $\hat{\de}\Psi=0$. In
the following two subsections these two cases are examined
individually.

\subsection{Calculation of the Interface  Matrix}
\label{matrices}

We assume that the refractive index changes in a step-like fashion
along a nonstationary interface in the $(z,ct)$-plane, postponing
the discussion of the validity of this sharpness assumption to the
end of the next paragraph. The generic form of the refractive index
is then given by
$n(z,t)=n_{1}+(n_{2}-n_{1})\Theta(\cos(\theta)z-\sin(\theta)ct)$,
where $n_{1,2}$ are the refractive indices of the first and the
second media, and $\Theta$ is the Heaviside function.  $\theta$
denotes the angle between the $ct$-axis and the interface in the
$(z,ct)$-plane, which we will occasionally also parameterize via the
slope parameter $\beta=\tan(\theta)$. Note that the quantity $\beta
c$ represents the velocity of the nonstationary interface. This
velocity is arbitrary, and in particular it can be greater than the
local speed of light. This is not in conflict with special
relativity, because the nonstationary interface is not associated
with a moving medium, and it constitutes one of the main differences
between the present work and previous literature on nonstationary
dielectrics. For different aspects of the physics of moving media
the reader is referred to chapter IX of Landau \cite{landau},
section I.5 of Jackson \cite{jackson}, and to more recent works by
Yeh and collaborators \cite{yeh}, Saca \cite{saca}, Huang
\cite{huang} and by Leonhardt and Piwnicki \cite{leo}.

Our task is now to find an appropriate transfer matrix connecting
the amplitudes of the fields $F^{\pm}$ on either side of the
interface. We call this an {\em interface matrix}. Using the
appropriate orthogonal coordinates
\begin{eqnarray}
  \label{eq:pdef}
p(z,t)&=&\cos(\theta)z-\sin(\theta)ct = \frac{z-\beta ct}{\sqrt{1+\beta^2}},\\
\label{eq:qdef} q(z,t)&=&\sin(\theta)z+\cos(\theta)ct = \frac{\beta
z+ct}{\sqrt{1+\beta^2}},
\end{eqnarray}
normal ($p$) and parallel ($q$) to the interface in the
$(z,ct)$-plane, we can express the derivative operator in
Eq.(\ref{main}) as $\hat{\de}=\hat{A}\de_{p}+\hat{B}\de_{q}$, with
the matrices $\hat{A}$ and $\hat{B}$ defined as
\begin{eqnarray}
  \label{eq:2}
  \hat{A}&\equiv&[\sigma_{z}\cos(\theta)-n\sigma_{0}\sin(\theta)]\\
  \hat{B}&\equiv&[\sigma_{z}\sin(\theta)+n\sigma_{0}\cos(\theta)].\nonumber
\end{eqnarray}
Taking into account that $\de_{q} n = 0$ we can write
Eq.(\ref{main}) in the form
\begin{equation}
  \label{eq:3}
   \left[ \de_{p}  + \hat{A}^{-1}\hat{B}\de_{q} + \hat{A}^{-1} \sx
     \hat{A} (\de_{p} \ln{n})\sxplus  \right]\Psi=0.
\end{equation}
Writing Eq.\ref{eq:3} in the basis $[nE,B]^T=(\sx + \sz)\Psi$ yields
the following set of ordinary differential equations:
\begin{eqnarray}
  \label{eq:16}
  \partial_{p}\left(\ln nE\right)+\frac{1}{n^{2}\beta^{2}-1}
  \left[-\beta\left(1+n^{2}\right)\partial_{q}(\ln nE)\right. & & \\
  \left. \nonumber -n\left(1+\beta^{2}\right)
    \frac{\partial_{q}B}{\left(nE\right)}+
    \left(1+n^{2}\beta^{2}\right)\left(\partial_{p}n\right)\right]
  &=&0,\\
\label{eq:17}
  \partial_{p}(B)+\frac{1}{n^{2}\beta^{2}-1}
  \left[- \beta\left(1+n^{2}\right)\partial_{q}(B)\right. & & \\
  \left. \nonumber -n\left(1+\beta^{2}\right)\partial_{q}\left(nE\right) +
    2n\beta\left(\partial_{p}n\right)\left(nE\right)\right]&=&0.
\end{eqnarray}
We can now integrate Eq.~(\ref{eq:16}) from $p=-\epsilon$ to
$p=+\epsilon$ across the interface at $p=0$, where $\eps$ is an
arbitrarily small positive number. If $B$ and $(nE)$ are
differentiable with respect to $q$ (i.e. $\partial_qB<\infty$ and
$\partial_q \ln nE <\infty$), and additionally $n^2\beta^2 \neq 1$
across the interface, then the contributions to this integral from
terms containing $\partial_{q}(B)$ and $\partial_{q}\left(nE\right)$
are proportional to $\epsilon$ and will therefore vanish in the
limit $\epsilon \to 0$; Eq.~(\ref{eq:16}) can then be integrated
analytically. Under the same conditions Eq.~(\ref{eq:17}) can also
be integrated analytically and we obtain a linear relation between
$[n_1E_1,B_1]^T$ and $[n_2E_2,B_2]^T$. In the $[F^+,F^-]^T$ basis
this relation can be expressed via the interface matrix $D^{[12]}$
for the step-like transition that connects the fields $F^{\pm}$ in
medium 1 with those in medium 2: \eq{deftransf}{\left(
\begin{array}{c} F^{+}_{1}
\\ F^{-}_{1}
\end{array}\right)=D^{[12]}\left( \begin{array}{c} F^{+}_{2} \\ F^{-}_{2}
\end{array}\right).} Explicitly we find
\eq{transfer}{D^{[12]}=\frac{1}{2n_{2}}\left(
\begin{array}{cc} (n_{1}+n_{2})\frac{1-n_{2}\beta}{1-n_{1}\beta}, &
(n_{1}-n_{2})\frac{1+n_{2}\beta}{1-n_{1}\beta} \\
(n_{1}-n_{2})\frac{1-n_{2}\beta}{1+n_{1}\beta}, &
(n_{1}+n_{2})\frac{1+n_{2}\beta}{1+n_{1}\beta}
\end{array}\right).} This interface matrix  possesses the classical property
$(D^{[12]})^{-1}=D^{[21]}$. For future reference it is also useful
to calculate
\begin{equation}
  \label{eq:7}
  \det
  D^{[12]}=\left(\frac{1-n_2^2\beta^2}{1-n_1^2\beta^2}\right)\frac{n_{1}}{n_{2}}.
\end{equation}

Note that in the derivation of \ref{transfer} we have used that
$n^2\beta^2 \neq 1$ as we cross the interface. In particular this
implies that our derivation is not valid in the regime where
$1/n_2<|\beta|<1/n_1$. The physical reason for this will be
discussed at the end of the next subsection, after we introduce the
free propagation matrix.

\subsection{Free Propagation in Real Space}
\label{sec:calc-prop-matr}

From the knowledge of the general transfer matrix for a
nonstationary interface [Eq.(\ref{transfer})], it is possible to
construct the transfer matrix for more complicated nonstationary
objects, such as multilayer structures, which will be done in
Section \ref{applications}. However, we still need to construct one
missing ingredient, the so-called {\em propagation matrix}, which
models the propagation of the fields along the $p$-direction far
from the interfaces, where the refractive index is constant. Under
this condition, Eq.(\ref{main}) simply becomes $\hat{\de}\Psi=0$,
which using the definitions of Eq.~\ref{eq:2} can be expressed as
\eq{eqforpsi}{\hat{A}\de_{p}\Psi+\hat{B}\de_{q}\Psi=0.} To solve
Eq.(\ref{eqforpsi}), we first write the derivatives $\de_{p,q}$ as
first order finite differences, namely $\de_{p}\Psi=[\Psi(p+\Delta
p,q)-\Psi(p,q)]/\Delta p$ and $\de_{q}\Psi=[\Psi(p,q+\Delta
q)-\Psi(p,q)]/\Delta q$, and then substitute into the equation:
\eq{finitediff1}{\hat{A}\frac{\Psi(p+\Delta p,q)-\Psi(p,q)}{\Delta
p}+\hat{B}\frac{\Psi(p,q+\Delta q)-\Psi(p,q)}{\Delta q}=0.} For a
fixed $\Delta p$, we can eliminate the term proportional to
$\Psi(p,q)$ in Eq.(\ref{finitediff1}) with an appropriate choice of
$\Delta q$, different for the two components of the $\Psi$-vector.
From Eqs.\ref{eq:2} one obtains \eq{finitediff2}{\Delta
q^{\pm}=\frac{\sin(\theta)\pm n\cos(\theta)}{\cos(\theta)\mp
n\sin(\theta)}\Delta p,} where the refractive index $n$ refers to
the homogeneous layer under consideration, and $\Delta q^{\pm}$
refers to the component $F^{\pm}$. With the particular choice of
Eq.(\ref{finitediff2}), the free propagation of the optical fields
for constant refractive index assumes a particularly simple and
effective form, which we write for the two components:
\eq{propagator}{F^{\pm}(p+\Delta p,q)=F^{\pm}(p,q+\Delta q^{\pm}).}
This means that advancing the field $F^{\pm}$ along the
$p$-direction by a small step $\Delta p$ (keeping $q$ constant), is
equivalent to advancing the same field by an amount equal to $\Delta
q^{\pm}$ along the $q$-direction (keeping $p$ constant). The
geometrical interpretation of Eq.(\ref{propagator}) is that the
points $(p+\Delta p,q)$ and $(p,q+\Delta q^{\pm})$ belong to the
same light cone.

The concept of the interface matrix [Eq.(\ref{transfer})] expressed
in Section \ref{matrices} is only valid for a sufficiently sharp
interface, i.e. if the interface transition layer is sufficiently
thin, such that the effect of the propagator (\ref{propagator}) can
be neglected while traversing the interface. For a given interface
width $\Delta p$ we can calculate via (\ref{finitediff2}) the length
of the interval $\Delta q^{\pm}$ along the interface which is
relevant for the propagator. If $F^{\pm}$ does not change
significantly on this interval, i.e. if
\begin{equation}
\Big{|}\de_q \ln F^{\pm} \frac{\beta \pm n}{1 \mp n\beta}\Delta
p\Big{|} \ll 1, \label{eq:5}
\end{equation}
then the propagator is negligible and the use of the interface
matrix of Eq.(\ref{transfer}) is justified. Note however that the
left hand side of Eq.\ref{eq:5} becomes singular for $n\beta = \pm 1
$, and then \ref{eq:5} will not be fulfilled for any choice of
$\Delta p$. In particular, for $n_{2}>n_{1}$, if $1/n_2 < \beta <
1/n_1$, then during the integration of \ref{eq:3} we will reach the
point where the singularity in \ref{eq:5} occurs. This is precisely
at the point where the matrix $\hat{A}^{-1}$ also becomes singular.
The condition on the sharpness of the interface which was used in
the derivation of Eq.(\ref{transfer}) can therefore in principle not
be fulfilled and therefore we shall not use Eq.(\ref{transfer}) in
the regime of $1/n_2 < \beta < 1/n_1$.

\subsection{Plane wave expansion. Calculation of Propagation Matrix}
\label{sec:plane-wave-expansion}

So far we have expressed the interface matrix of Eq.(\ref{transfer})
and the free propagator of Eq.(\ref{propagator}) as operating on the
fields vector $\Psi(p,q)$, where $p$ and $q$ are the rotated
variables of Eqs.~(\ref{eq:pdef}) and (\ref{eq:qdef}). However, the
interface matrix of Eq.(\ref{transfer}) connects the fields $F^{+}$
and $F^{-}$ only along the $p$-direction, while not affecting the
fields along the $q$-direction at all. It is therefore natural to
consider a plane wave expansion along the $q$-axis,
\begin{equation}
  \label{eq:9}
  \Psi(p,q) = \sum_{\tilde{\omega}}  \Psi^{\tilde{\omega}}(p)
  e^{-i\tilde{\omega}q}.
\end{equation}
Then Eq.(\ref{deftransf}) transforms into
\begin{equation}
\Psi^{\tilde{\omega}}_1 = D^{[12]}
\Psi^{\tilde{\omega}}_2,\label{eq:1}
\end{equation}
which means that the generalized frequency $\tilde{\omega}$ is
conserved across the interface. $\tilde{\omega}$ has the dimension
of a wavenumber, and physically it represents the momentum
associated to the propagation along the $q$-direction.

Inside a homogeneous medium we can expand $\Psi(p,q)$ in plane waves
along both the $p$ and $q$ axes:
\begin{equation}
\label{eq:11}
  \Psi(p,q) = \sum_{\tilde{\omega},\tilde{k}}
  \Psi^{\tilde{\omega},\tilde{k}} e^{i(\tilde{k}p-\tilde{\omega}q).}
\end{equation}
Since each plane wave component has to fulfill the propagator
equation $\hat{\de}\Psi^{\tilde{\omega},\tilde{k}}
e^{i(\tilde{k}p-\tilde{\omega}q)}=0$, we obtain for the two
components $F^{\pm}$ the following dispersion relations:
\begin{equation}
  \label{relation}
  \tilde{k}^{\pm}=\frac{\sin(\theta)\pm
n\cos(\theta)}{\cos(\theta)\mp n\sin(\theta)}\tilde{\omega},
\end{equation} which can also be directly deduced from Eq.(\ref{finitediff2}). This
enables us to rewrite the free propagator (\ref{propagator}) for the
Fourier components as
\begin{equation}
  \label{eq:8}
\Psi^{\tilde{\omega}}(p+\Delta p)  = P(\Delta p)
\Psi^{\tilde{\omega}}(p)= \left(
\begin{array}{cc} e^{i\tilde{k}^{+}\Delta p} & 0 \\ 0 &
e^{i\tilde{k}^{-}\Delta p}
\end{array} \right)\Psi^{\tilde{\omega}}(p),
\end{equation}
an expression that will be useful in Section
\ref{sec:photoniccrystals}.

As a final issue, it is worth writing explicitly the transformation
law between the momenta ($\tilde{k}$,$\tilde{\omega}$), reciprocal
to the real-space variables $(p,q)$ respectively, and $(k,\omega)$,
reciprocal to the variables $(z,ct)$. By using
Eqs.(\ref{eq:pdef})-(\ref{eq:qdef}) and the expansion of
Eq.(\ref{eq:11}) we obtain: \eq{rotation}{\left(
\begin{array}{c} \tilde{k} \\ \tilde{\omega} \end{array}
\right)=\left(
\begin{array}{cc} \cos(\theta) & \sin(\theta) \\ -\sin(\theta) & \cos(\theta) \end{array} \right)\left(
\begin{array}{c} k \\ \omega/c \end{array} \right).} From the simple
rotation of Eq.(\ref{rotation}) one can deduce some important
physical consequences. First of all for $\theta=0$ ($\beta=0$) it
follows that $\tilde{k}=k$ and $\tilde{\omega}=\omega/c$, so that in
the static case the $p$-direction coincides with the $z$-axis and
the $q$-direction coincides with the $ct$-axis by definition. But
because the $q$-direction is associated with the delocalization
direction of an arbitrary plane wave in Eq.(\ref{eq:11}), it follows
that during the scattering process of light by a static interface
the quantity $\tilde{\omega}=\omega/c$ is conserved. This
corresponds to the energy conservation at the interaction point. In
other words, all generated waves will have exactly the same
frequency. The other limiting case is when $\theta\rightarrow\pi/2$
($\beta\rightarrow\infty$). For this case the role of frequency and
wavenumber is reversed, and from Eq.(\ref{rotation}) we have
$\tilde{k}=\omega/c$ and $\tilde{\omega}=-k$. The delocalization
direction is still along $q$, which now points along the $z$-axis.
Therefore during the scattering it is the wavenumber (i.e. the
momentum) that is conserved at the interaction point, and in general
all generated waves will have different frequencies but the same
momentum. For an intermediate situation ($0<\theta<\pi/2$), the
conserved quantity $\tilde{\omega}$ corresponds to a combination of
$k$ and $\omega/c$ according to the rotation of Eq.(\ref{rotation}).

\section{Applications}
\label{applications}

In the present section we discuss three fundamental applications of
the concepts expressed in Section \ref{matrices}, and in particular
of the transfer interface matrix given by Eq.(\ref{transfer}) and
the free propagator given in real space by Eq.(\ref{propagator}) and
in Fourier space by Eq.(\ref{eq:8}).

In subsection \ref{nonstationaryinterface} we analyze the physics of
scattering by nonstationary interfaces in different interesting
physical configurations, shown in Fig. \ref{fig2}. In subsection
\ref{sec:photoniccrystals} we calculate the bandstructure and the
forbidden bandgaps of a nonstationary photonic crystal, finding a
novel effect: the closure of the frequency bandgap for a certain
value of $\theta=\theta_{cr1}$, and the consequent opening of the
wavenumber bandgap for another $\theta=\theta_{cr2}$. In subsection
\ref{spacetimelenses} we show that it is possible to think of a
dynamical device, analogous to a spatiotemporal lens, that
efficiently performs pulse broadening and compression, with
simultaneous frequency/wavenumber manipulation of pulses.

\subsection{Nonstationary interfaces}
\label{nonstationaryinterface}

\begin{figure}
\includegraphics{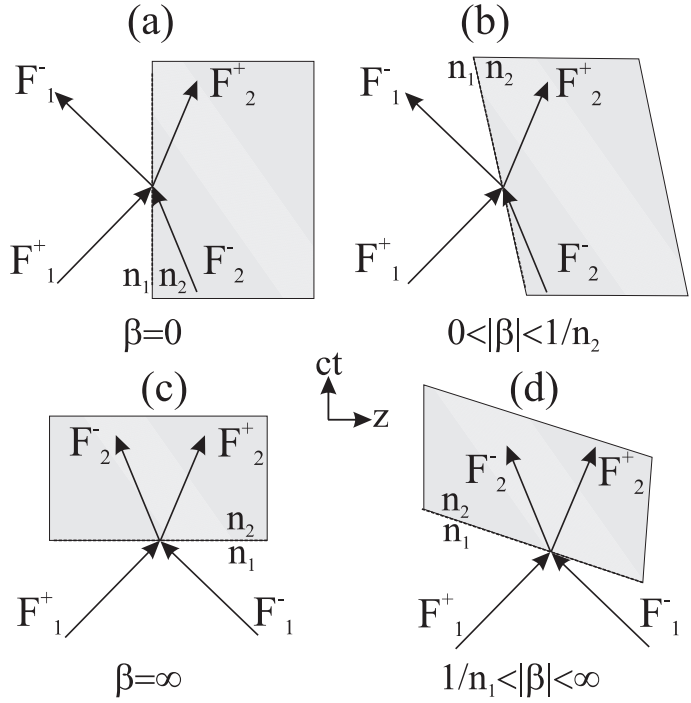}
\caption{\label{fig2} Panel showing space-time diagrams of
scattering of forward and backward fields $F^{\pm}_{j}$ ($j=\{ 1,2
\}$) from the four most important interface configurations. (a)
Space-like interface ($\beta=0$, $\theta=0$), which coincides with
the conventional, static interface.  (b) Interface configuration
with $\beta<0$ and $0<|\beta|<1/n_{2}$. (c) Time-like interface,
$\beta=\pm\infty$ ($\theta=\pm\pi/2$). (d) Interface configuration
with $\beta<0$ and $1/n_{1}<|\beta|<\infty$.}
\end{figure}

\begin{figure}
\includegraphics{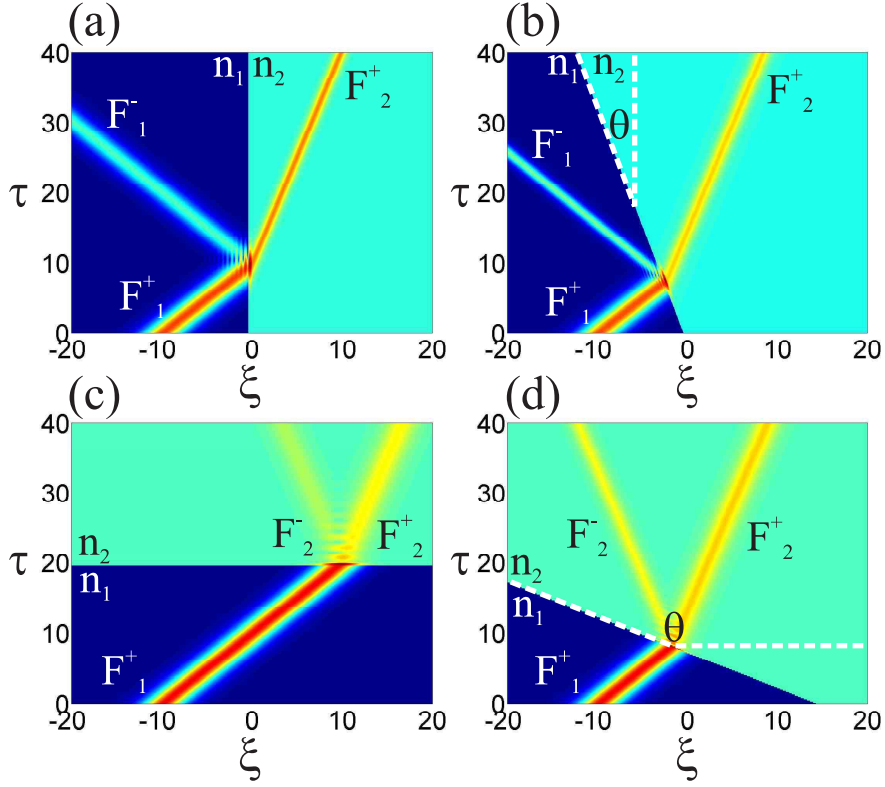}
\caption{\label{interfacesim} (Color online) Panel showing the
simulation of interaction of an incoming pulse with the four sharp
interface configurations of Fig. \ref{fig2}. (a) Space-like
interface: $\beta=0$. (b) $\beta=-0.3$. Angle $\theta=-0.2915$
radians is indicated with dashed lines. (c) Time-like interface:
$\beta=\infty$. (d) $\beta=-2$. Angle $\theta=-1.1071$ radians is
indicated with dashed lines. Refractive indices are: $n_{1}=1$ and
$n_{2}=3$. The agreement with theoretical expressions is excellent
for all cases. Input pulse parameter $w$ in Eq.(\ref{pulseform}) is
$w=1$. Horizontal axis is dimensionless space coordinate
$\xi=z/\lambda$, while the vertical axis is dimensionless time
coordinate $\tau=ct/\lambda$, and $\lambda$ is the central pulse
wavelength.}
\end{figure}

The panel of Fig.~\ref{fig2} shows the four most important cases of
interface configurations which can be analyzed with the help of
Eq.(\ref{transfer}). Let us first consider Fig. \ref{fig2}(a) which
shows the space-time diagram of a conventional static interface
($\theta=0$, $\beta=0$) between two media of refractive indices
$n_{1}$ and $n_{2}$ respectively in the plane $(z,ct)$. Because all
the variation of the refractive index is spatial, and the variable
$p$ defined in Eq.~(\ref{eq:pdef}) can be identified with  the
spatial coordinate $z$, we call this a {\em space-like} interface.
By taking the limit $\beta\rightarrow 0$ in Eq.(\ref{transfer}) we
find the well-known result for the interface matrix:
\eq{interfacea}{D^{[12]}_{\beta=0}=\frac{1}{2n_{2}}\left(
\begin{array}{cc} n_{1}+n_{2}, & n_{1}-n_{2}
\\ n_{1}-n_{2}, & n_{1}+n_{2} \end{array} \right).} The  transmission
and reflection coefficients are then classically defined for
$F_2^{-}=0$ as
\begin{eqnarray}
  \label{eq:t1}
  t&\equiv& \frac{F^{+}_2}{F_1^{+}} = \frac{ 1}{ D^{[12]}_{11}} =
  \frac{2n_2}{n_1 + n_2},\\
  \label{eq:r1}
  r&\equiv& \frac{F^{-}_1}{F_1^{+}} = \frac{ D^{[12]}_{21}}{ D^{[12]}_{11}} =
  \frac{n_1-n_2}{n_1 + n_2}.
\end{eqnarray}
For $0<|\beta|< 1/n_2$ as in Fig.~\ref{fig2}(b) the configuration is
similar to the one in Fig.~\ref{fig2}(a). From  the interface matrix
Eq.(\ref{transfer}) we can now calculate the generalized
transmission and reflection coefficients according to the
definitions of Eqs.(\ref{eq:t1})-(\ref{eq:r1}) as
\begin{equation}
  \label{eq:4}
  t= \frac{2n_2}{n_1 + n_2} \frac{1-n_1\beta}{1-n_2\beta},\quad
  r=\frac{n_1-n_2}{n_1 + n_2} \frac{1-n_1\beta}{1+n_1\beta}.
\end{equation}
Note that for $\beta<0$ and inside the parameter region $0<|\beta|<
1/n_2$, the expressions given by Eq.(\ref{eq:4}) never show any
singularity in the denominator. However, for $\beta>0$, we have a
singularity in the transmission coefficient for $\beta\rightarrow
1/n_{2}$. This limit corresponds to the disappearance from the
equations of one of the waves participating in the scattering
($F^{+}_{2}$), which will travel exactly along the interface, so
that $t$ as given by Eq.(\ref{eq:4}) cannot be used, and for
$\beta>1/n_{2}$ the incident light never reaches the interface.

For $|\beta|=\infty$ Fig. \ref{fig2}(c) shows the configuration when
the refractive index changes abruptly and simultaneously in time
throughout all space (for some discussion of the experimental
relevance of this concept see the conclusions, Section
\ref{conclusions}). Because the variation of the refractive index
occurs only along the $ct$-axis, we shall call this a {\em
time-like} interface. From Eq.~(\ref{transfer}) we obtain the
interface matrix
\eq{interfacec}{D^{[12]}_{\beta=\infty}=\frac{1}{2n_{1}}\left(
\begin{array}{cc} n_{1}+n_{2}, & n_{2}-n_{1}
\\ n_{2}-n_{1}, & n_{1}+n_{2} \end{array} \right).} For the
calculation of the transmission and reflection coefficients we have
now to set $F_1^{-}=0$, and then instead of Eqs.~(\ref{eq:t1}) and
(\ref{eq:r1}) we obtain
\begin{eqnarray}
  \label{eq:t2}
  t&\equiv& \frac{F^{+}_2}{F_1^{+}} = D^{[21]}_{11} =
  \frac{n_1 + n_2}{2n_2},\\
  \label{eq:r2}
  r&\equiv& \frac{F^{-}_2}{F_1^{+}} = D^{[21]}_{21}  =
   \frac{n_1 - n_2}{2n_2}.
\end{eqnarray}
Similar formulas for this special case have been given by
Morgenthaler in 1958 \cite{morgenthaler}. For a more general
$|\beta| > 1/n_1$ when the interface is tilted with respect to the
time-like case, the configuration is depicted in Fig.~\ref{fig2}(d).
Then Eqs.~(\ref{eq:t2}) and (\ref{eq:r2}) generalize to
\begin{eqnarray}
  \label{eq:6}
  t = \frac{1-\beta n_1}{1- \beta n_2}\frac{n_1 + n_2}{2n_1}, \quad r
  = \frac{1-\beta n_1}{1+ \beta n_2}\frac{n_2 - n_1}{2n_1}.
\end{eqnarray}
Even in this case, the expressions of $t$ and $r$ in Eq.(\ref{eq:6})
have singularities for $\beta=1/n_{2}$ and $\beta=-1/n_{2}$
respectively, which however are outside the region of validity of
these formulas, that is $|\beta|>1/n_{1}$.

In Section \ref{sec:plane-wave-expansion} we have seen that the
frequency $\tilde{\omega}$ arising in the plane wave expansion along
$q$ in Eq.(\ref{eq:1}) is conserved across the interface. As we have
already discussed previously, in the case of $\beta=0$
[Fig.~\ref{fig2}(a)] this is, with $\tilde{\omega}=\omega/c$, the
well-know frequency (i.e. energy) conservation at a spatial
interface. In the more interesting case $\beta=\infty$
[Fig.~\ref{fig2}(c)], however, we see that $\tilde{\omega}=-k$,
which means that wavenumber (i.e. momentum) is conserved at a
time-like interface.

The scattering of finite pulses from the interfaces for all four
cases shown in Fig. \ref{fig2} have been simulated by solving
numerically Eq.(\ref{main}), and results are shown in the panel of
Fig. \ref{interfacesim}. Electric field for the input pulse is taken
of the form \eq{pulseform}{E_{ini}\equiv E(z,t=0)=E_{0}\ \sech{
\frac{z}{w\lambda}}\cos\left(\frac{2\pi z}{\lambda}\right),} where
$E_{0}$ is the incident amplitude, and $w$ is a dimensionless
parameter which measures the spatial width in units of $\lambda$,
the pulse central wavelength. Without loss of generality, the
incident amplitude is in all cases normalized to the value
$E_{0}=1$.

We have used a second order finite difference algorithm to model the
derivatives in Eq.(\ref{main}) for the two components, and a fourth
order Runge-Kutta algorithm to advance the field in time. Once the
spatial distribution of the electric field over the $z$-axis at the
initial instant of time $E_{ini}=E(z,t=0)$ is given, the initial
magnetic field $B_{ini}=B(z,t=0)$ can be found by imposing that
there is no backward wave propagating for $t=0$, i.e. $F^{-}=0$,
obtaining $B_{ini}(z)=n(z,t=0)E_{ini}(z)$, where $n(z,t=0)$ is the
initial refractive index distribution. We use Neumann-type boundary
conditions, which means that we impose that the derivatives of the
optical fields vanish at the boundaries of the spatial window, while
the evolution of the fields is calculated by advancing in time. In
order to ensure complete stability of the numerical algorithm it is
sufficient to impose that all fields are spatially localized and
that the temporal step of the grid $\Delta t$ and the spatial step
of the grid $\Delta z$ satisfy the following condition: $\Delta
t\ll\Delta z^{2}$. For our simulations we always use a spatial
sampling of the optical fields with a number of points equal to
$N=2^{15}$, which is more than sufficient in order to obtain
extremely accurate results for long propagation times, and to avoid
the dangers of artifacts due to numerical dispersion during the
temporal evolution.

The panel of Fig. \ref{interfacesim} shows propagation of the
envelope of the electric field (with the fast oscillations decoupled
for sake of clarity) during the scattering processes analogous to
Figs. \ref{fig2}(a,b,c,d) respectively. Here and in the following,
contour plots are given in the dimensionless variables
$\tau=ct/\lambda$ and $\xi=z/\lambda$, with $\lambda=2\pi
c/\omega_{0}$, and $\omega_{0}$ is the central circular frequency of
the initial pulse. Note that Eq.(\ref{main}) is solved directly in
dimensionless units, therefore we do not need to specify the scaling
wavelength $\lambda$, and results shown in the figures can be
adapted to the particular spatial scale that one chooses to
consider. The extension of the spatial grid used is
$\xi_{window}=300$ in dimensionless units, which is relatively large
in order to ensure total spatial localization of all waves
participating in the scattering, although only a small portion of
this window is shown in the figures. The spatiotemporal steps are
$\Delta\xi=9.15\times 10^{-3}$ and $\Delta\tau=7.67\times 10^{-7}$.
The pulse is initially propagating at the speed of light in vacuum
($n_{1}=1$). The sharp interface is modeled through a Heaviside
function $n(z,t)=n_{1}+(n_{2}-n_{1})\Theta(z-\beta ct)$, with
$n_{2}=3$. The values of $\beta$ for Fig. \ref{fig2}(a,b,c,d) are
respectively $\beta=0$, $\beta=-0.3$, $\beta=\infty$, and
$\beta=-2$. In all cases, $t$ and $r$ as given by the theoretical
analysis given above are predicted with a high degree of accuracy.

\begin{figure}
  \includegraphics{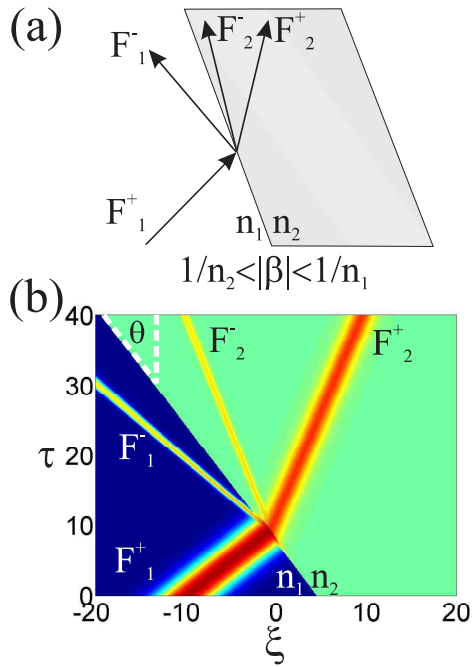}
  \caption{(Color online) (a) Scattering diagram in the case $\beta<0$, $1/n_2 < |\beta| < 1/n_1$,
  and (b) corresponding numerical simulation. Parameters are: $\beta=-0.4$ ($\theta=-0.3805$ radians, indicated
  with dashed lines), $w=1$. Refractive indices are: $n_{1}=1$ and $n_{2}=3$.}
  \label{fig:paradox}
\end{figure}

As discussed at the end of Section \ref{sec:calc-prop-matr}, the
derivation of the interface matrix (\ref{transfer}) does not cover
the case of $1/n_2 < |\beta| < 1/n_1$. The configuration for this
case is depicted in the upper panel of Fig.~\ref{fig:paradox}. We
see that this case resembles the space-like configuration as in
Fig.~\ref{fig2}(a) in the region $n=n_1$ and the time-like
configuration as in Fig.~\ref{fig2}(c) for $n=n_2$. Note that now
only one of the four fields ($F_1^{+}$) reaches the interaction
point from the past, while the other three fields propagate from the
interface in positive time direction. It is therefore not clear how
in this case the transmission or reflection coefficient can be
consistently defined, and neither of the given definitions of
Eqs.~(\ref{eq:t1})-(\ref{eq:r1}) nor
Eqs.~(\ref{eq:t2})-(\ref{eq:r2}) are satisfactory in this case. The
theoretical origin of this problem is that in the region $1/n_2 <
|\beta| < 1/n_1$ the transfer matrix concept is not applicable, due
to the fact that for these values of $\beta$ the concept of sharp
interface cannot be defined, at least using a dispersionless
susceptibility, which does not make distinctions between long and
short wavelengths. In particular, the condition of Eq.(\ref{eq:5})
does not necessarily hold in the above region, as recognized by
Ostrovsky \cite{ostrovsky}. Nevertheless we can still solve
Eq.(\ref{main}) numerically as shown in Fig.~\ref{fig:paradox}(b),
and we leave the investigation of this complicated issue for a
future publication.

\subsection{Nonstationary Photonic Crystals and Bandgap Calculation. Spatiotemporal Bragg reflector}
\label{sec:photoniccrystals}

We now use the knowledge built in Section
\ref{sec:plane-wave-expansion} to calculate the bandstructure and
the forbidden bandgaps of a nonstationary multilayer structure for
different values of $\beta$, and in particular we shall enunciate
the Bragg condition for a generalized Bragg reflector.

\begin{figure}
\includegraphics{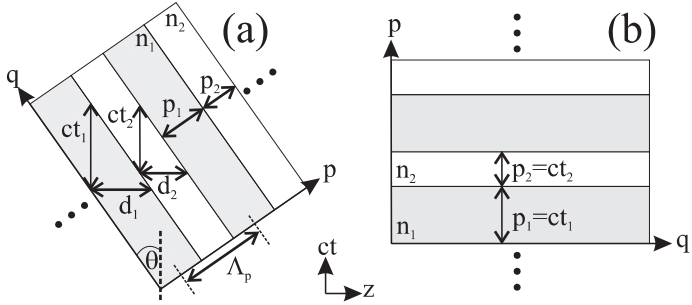}
\caption{\label{phcrgeometry} (a) Schematic figure showing the
quantities of interest used in the text to define a general
multilayer structure (with indices $n_{j}$, $j=\{ 1,2 \}$) for an
arbitrary angle $\theta$ (i.e. arbitrary parameter $\beta$). $p$
indicates the propagation direction for the Bloch waves, while $q$
is the delocalization direction for plane waves. $d_{j}$ are the
widths of the layers along the $z$-direction, while $ct_{j}$ are the
widths along the $ct$-direction. $p_{j}$ indicate the same
quantities but along the $p$-direction. $\Lambda_{p}=p_{1}+p_{2}$ is
the total length of the unit cell, also indicated in the figure. (b)
Time-like photonic crystal. The refractive index varies only along
the time axis, which therefore coincides with $p$. The coordinate
$z$ coincides with $q$, therefore plane waves are spatially
delocalized. For the structure to be a time-like Bragg reflector,
which is totally reflective for a certain wavenumber bandgap, the
condition $ct_{j}=n_{j}\lambda/4$ must hold, where $\lambda$ is the
wavelength of the incident plane wave.}
\end{figure}

Let us assume in the following analysis that the periodicity of the
multilayer structure is along the $p$-direction [see Fig.
\ref{phcrgeometry}(a) and caption for a description of the geometry
of the problem under consideration]. As in Section
\ref{sec:plane-wave-expansion} we now deal with the propagation of
plane waves of the form
$\Psi(p,q)=\Psi^{\tilde{\omega}}(p)e^{-i\tilde{\omega}q}$. Similarly
to the extensively studied Kronig-Penney model \cite{kronigpenney},
we now look for a matrix $T$ such that
\begin{equation}
  \label{eq:10}
  \Psi^{\tilde{\omega}}(p+\Lambda_p) = T \Psi^{\tilde{\omega}}(p)
\end{equation}
which propagates $\Psi^{\tilde{\omega}}(p)$ from the center of a
layer with refractive index $n_1$ across a layer with refractive
index $n_2$ into the center of the next layer with refractive index
$n_1$ as indicated in Fig.~\ref{phcrgeometry}(a). The length of this
unit cell is given by $\Lambda_{p}=p_{1}+p_{2}$, where $p_{j}$ are
the widths of the layers of indices $n_{j}$ along the $p$-direction.
From Eqs.~(\ref{eq:1}) and (\ref{eq:8}) we obtain
\begin{equation}
  \label{eq:12}
  T = P_1( p_1/2)D^{[12]}P_2( p_2)D^{[21]} P_1( p_1/2)
\end{equation}
where $P_j(p)$ is the propagation matrix according to
Eq.(\ref{eq:8}) in the medium with refractive index $n_j$. Since
$D^{[12]}D^{[21]} = 1$ we observe that $|\det T|=1$. Considering the
two eigenvalues $\lambda_1^T$ and  $\lambda_2^T$  of $T$ we can then
distinguish between the following two cases: (i)
$|\lambda_1^T|=|\lambda_2^T|=1$ and (ii) $|\lambda_1^T|>1$ and
$|\lambda_2^T|<1$.

\begin{figure}
\includegraphics{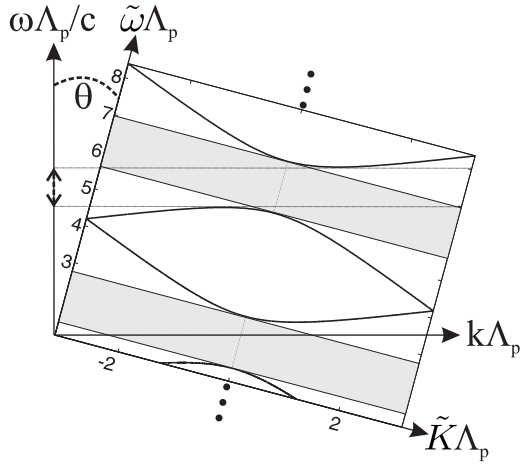}
\caption{\label{bandstructure} Bandstructure (solid lines) and
bandgaps (grey areas) of a spatiotemporal Bragg reflector that
satisfies the generalized quarter wavelength condition of
Eq.(\ref{quarterwave}), as calculated by using the transcendental
photonic band equation Eq.(\ref{eq:15}). According to
Eq.(\ref{rotation}), the orthogonal axes $\tilde{\omega}$ and
$\tilde{k}$ are rotated with respect to the orthogonal axes $\omega$
and $k$ by an angle $\theta$, as indicated in the figure. All
quantities shown are dimensionless, and normalized in units of
$\Lambda_{p}$. For $\theta=0$ (space-like photonic crystal), the two
orthogonal bases would coincide, i.e. $\tilde{\omega}=\omega/c$ and
$\tilde{k}=k$, and this corresponds to the case of a conventional,
static photonic crystal. For a general $\theta$, the forbidden
bandgaps in $\tilde{\omega}$ as calculated by Eq.(\ref{eq:15}) are
shown by the grey areas. The projection of one of the bandgaps onto
the axis $\omega$ is also shown, with its new smaller extension
indicated by a dashed double-arrow. In the limit
$\theta\rightarrow\pi/2$ (time-like photonic crystal),
$\tilde{\omega}$ tends to $-k$, while $\tilde{k}$ tends to
$\omega/c$, so that the forbidden gaps in $\tilde{\omega}$ will
correspond to bandgaps in $k$. Note that there is a critical angle
$\theta=\theta_{cr1}$ at which the projected $\omega$-gap will
close, and a second critical angle $\theta_{cr2}>\theta_{cr1}$ at
which the $k$-bandgaps start to open, until they reach their full
extension at $\theta=\pi/2$.}
\end{figure}

Free propagation will only occur in case (i), which by elementary
algebra can be shown to be equivalent to
\begin{equation}
  \label{eq:13}
  \frac{1}{2}|\tr{T}| < 1.
\end{equation}
In this case it is convenient to write the eigenvalues of $T$ in the
form
\begin{equation}
  \label{eq:18}
  \lambda_1=e^{i\bar{k}\Lambda_p}e^{i\tilde{K}\Lambda_p} \qquad
  \lambda_2=e^{i\bar{k}\Lambda_p}e^{-i\tilde{K}\Lambda_p},
\end{equation}
where $\bar{k}$ is calculated via $\sqrt{\det
T}=e^{i\bar{k}\Lambda_p}$ and $\tilde{K}$ fulfills the relation
\begin{equation}
  \label{eq:14}
  \cos(\tilde{K}\Lambda_{p})=\frac{\tr{T}}{2 \sqrt{\det T}}=\frac{
    \lambda_1 +  \lambda_2}{2 e^{i\bar{k}\Lambda_p}}.
\end{equation}
Physically $\bar{k}$ simply gives rise to a common phase factor in
both eigenvalues of $T$, while $\tilde{K}$ can be interpreted as a
generalization of the Bloch wavenumber in the Kronig-Penney model
for nonstationary interfaces. In particular the explicit calculation
of $\tilde{K}$ via a straightforward analytic expansion of the trace
and determinant of $T$ yields the familiar-looking expression
\cite{kronigpenney}:
\begin{eqnarray}
  \label{eq:15}
  \cos(\tilde{K}\Lambda_{p})&=\cos(p_{1}\tilde{k}^\Delta_1)
  \cos(p_{2}\tilde{k}^\Delta_2)\\
  & \nonumber
  - \frac{n_{1}^{2}+n_{2}^{2}}{2n_{1}n_{2}}
  \sin(p_{1}\tilde{k}^\Delta_1)
  \sin(p_{2}\tilde{k}^\Delta_2),
\end{eqnarray}
where
\begin{equation}
  \label{transcendental2}
  \tilde{k}^\Delta_j = \frac{ \tilde{k}^+_j- \tilde{k}^-_j}{2}=
  \frac{n_j\tilde{\omega}}{\cos^2(\theta)-\sin^2(\theta)n_j^2}=
  \frac{(1+\beta^2)n_j\tilde{\omega}}{1-\beta^2n_j^2.}
\end{equation}
For a given $\tilde{\omega}$ we can now use
Eqs.~(\ref{eq:15})-(\ref{transcendental2}) to calculate the
appropriate $\tilde{K}$, if it exists. This gives rise to a band
structure relation as shown in Fig.~\ref{bandstructure} (see also
caption). Note that $\beta$ only enters in the band structure
calculation via $\tilde{k}^\Delta_j$, which however diverges for
$\beta \rightarrow 1/n_j$ [cf.\ Eq.~(\ref{transcendental2})].

The forbidden bandgaps are located in those regions of
$\tilde{\omega}$ for which condition (\ref{eq:13}) is not fulfilled,
i.e.\ the right hand side of Eq.~(\ref{transcendental2}) is larger
than unity and $\tilde{K}$ would have a non-vanishing imaginary
part, making the Bloch waves evanescent. In particular if the
quantities $p_{j}$ are chosen such that
\begin{equation} p_{j}\tilde{k}_j^{\Delta}=\pi/2,\label{quarterwave}
\end{equation}
then one can form a Bragg reflector with total reflectivity inside
the forbidden bandgaps in $\tilde{\omega}$. Condition
(\ref{quarterwave}) is analogous to the quarter wavelength condition
typical of a Bragg stack. However, and this is the crucial point, it
should be noted that the meaning of $\tilde{\omega}$ is different
for different values of $\beta$. For instance, in the traditional
case of space-like Bragg stack, i.e. in the limit $\beta\rightarrow
0$, condition (\ref{quarterwave}) becomes the well-known $\omega
d_{j}n_{j}/c=\pi/2$, because in this limit
$\tilde{k}_j^{\Delta}\rightarrow n_j \tilde{\omega}$,
$\tilde{\omega}\rightarrow\omega/c$ and $p_{j}\rightarrow d_{j}$
(the width of layer $j$), where $\omega$ is the initial plane wave
frequency. However, in the time-like Bragg stack, i.e. in the limit
$\beta\rightarrow\infty$ [see Fig. \ref{phcrgeometry}(b)],
Eq.(\ref{quarterwave}) becomes $kt_{j}/n_{j}=\pi/2$, because in this
limit $\tilde{k}_j^{\Delta}\rightarrow \tilde{\omega}/n_j$,
$\tilde{\omega}\rightarrow -k$, and $p_{j}\rightarrow ct_{j}$ (the
duration of layer $j$) where $k$ is the initial plane wave
wavenumber, common to all waves.

Fig. \ref{bandstructure} shows the bandstructure (solid lines) and
the forbidden bandgaps (grey areas) in $\tilde{\omega}$ for a
generalized Bragg reflector that satisfies the generalized quarter
wavelength condition of Eq.(\ref{quarterwave}), with
$0<\theta<\pi/2$, as calculated by solving numerically the
transcendental photonic band equation Eq.(\ref{eq:15}). For this
special case, all bandgaps have exactly the same extension and the
spacing between them is regular \cite{pochiyeh}. All momenta in Fig.
\ref{bandstructure} are normalized to dimensionless units by
multiplying by $\Lambda_{p}$, and $\tilde{K}$ is the Bloch
wavenumber. Orthogonal basis $(\tilde{\omega},\tilde{k})$ is rotated
with respect to the basis $(\omega/c,k)$ of an angle $\theta$,
indicated in the figure, see also Eq.(\ref{rotation}). If
$\theta\rightarrow 0$, $\tilde{\omega}\rightarrow\omega/c$, the two
axes coincide and the bandgaps in $\tilde{\omega}$ would correspond
exactly to the bandgaps in $\omega/c$. This is of course the
space-like case of a static, conventional Bragg reflector. For the
other extreme case, $\theta\rightarrow\pi/2$,
$\tilde{\omega}\rightarrow -k$, and the bandgaps in $\tilde{\omega}$
would correspond to bandgaps in $k$. This novel configuration
corresponds to a time-like Bragg reflector, of the type shown in
Fig. \ref{phcrgeometry}(b). In the intermediate cases
$0<\theta<\pi/2$, depicted in Fig. \ref{bandstructure}, the
projection of one of the bandgaps onto the $\omega$ axis is shown,
and its extension is marked with a dashed double-arrow. It is clear
that for increasing values of $\theta$ the extension of the
projected $\omega$-bandgap is shrinking, until for a certain
critical angle $\theta_{cr1}$ its size vanishes, indicating the
closure of  the frequency gap. A further increase of $\theta$ will
lead to a second critical angle $\theta_{cr2}$, in correspondence of
which the bandgaps projected onto the $k$-axis gradually start to
open, until they reach their full extension at $\theta=\pi/2$.

With the above discussions we have therefore identified the novel
general concept of a {\em spatiotemporal Bragg reflector}, and hence
of a {\em spatiotemporal photonic crystal} with a generalized
quarter wavelength condition given by Eq.(\ref{quarterwave}).

\subsection{Spatiotemporal Lenses, Pulse Compression and Broadening}
\label{spacetimelenses} 
\begin{figure}
\includegraphics{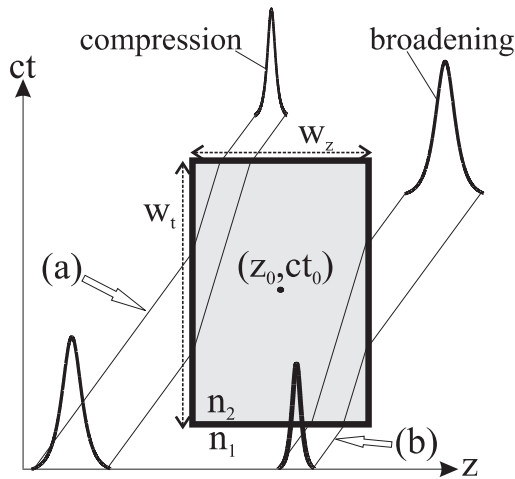}
\caption{\label{squarelens} Schematic figure showing the mechanism
of pulse compression [(a)] and pulse broadening [(b)] due to
spatiotemporal lensing. The refractive index of the rectangular
region shown is given by Eq. (\ref{indicelens}). The refractive
indices outside the rectangle and inside it are uniform, in the
$(z,ct)$-plane, and equal to $n_{1}$ and $n_{2}$ respectively. In
this example we assume that $n_{2}>n_{1}$. During the scattering
process every portion of the pulse will travel according to its
local light cone, which is narrower in the grey region (light
travels at speed $c/n_{2}<c/n_{1}$), and wider outside it. Backward
waves produced in the scattering are not indicated in the figure for
sake of clarity, but they are always present for sharp interfaces.}
\end{figure}

Another interesting and potentially useful
effect exhibited by a simple class of spatiotemporal dielectric
structures is what we have called {\em spatiotemporal lensing}. Let
us suppose that we have a refractive index distribution in the
$(z,ct)$-plane of the form:
\eq{indicelens}{n(z,t)=n_{1}+(n_{2}-n_{1})\exp\left[ -
\left(\frac{t-t_0}{w_{t}/2}\right)^{2m_{t}}
-\left(\frac{z-z_0}{w_{z}/2}\right)^{2m_{z}} \right].} In
Eq.(\ref{indicelens}), $z_{0}$ and $t_{0}$ represent the coordinates
of the center of the rectangular region, and $w_{z,t}$ are
respectively its spatial and temporal extension, see Fig.
\ref{squarelens}. The two integer numbers $m_{z,t}$ model the
sharpness of the supergaussian transitions between $n_{1}$ and
$n_{2}$ along the spatial and temporal directions respectively.
Eq.(\ref{indicelens}) models a rectangular region in the
$(z,ct)$-space where the transition between an 'external' refractive
index $n_{1}$ and an 'internal' index $n_{2}$ occurs, see Fig.
\ref{squarelens}. Here and in the following we shall refer to this
region as a {\em spatiotemporal lens}, for reasons that will be
clear shortly.

Fig. \ref{squarelens} illustrates the essential features of the
scattering of a generic pulse by the rectangular spatiotemporal lens
in the $(z,ct)$-plane. Essentially two configuration are of
interest. In the first configuration [Fig. \ref{squarelens}(a)],
after some free propagation in the medium with index $n_{1}$, the
pulse enters the grey region with index $n_{2}$ on the left side of
the rectangular region, with the entire body of the pulse well
inside the lateral extension of the rectangle ($w_{t}$). The pulse
then propagates inside the structure, at a greater angle in the
$(z,ct)$-plane, each portion of the pulse following the local light
cone (defined in Fig. \ref{geometry}) at any instant of propagation.
When the pulse exits the grey region, however, it is clear from Fig.
\ref{squarelens}(a) that its pulse duration has been reduced, and
pulse compression has been achieved. It can be shown by elementary
geometrical considerations that the compression ratio between the
output and the input pulses in this simple case is just
$\rho=n_1/n_2 $. The scaling in the $(z,ct)$-plane  is accompanied
by an increase of the central frequency of the pulse by $1/\rho$,
with a consequent increase of the spectral bandwidth. Thus we obtain
an effective method for manipulating the frequency and wavenumber of
a pulse.

The second scenario is when the pulse is delayed in such a way that
it enters the spatiotemporal structure on the bottom side of the
rectangle, the extension of which is $w_{z}$, and well inside it,
see Fig.  \ref{squarelens}(b). This case shows exactly the opposite
dynamics of the previous case, and pulse broadening in the
$(z,ct)$-plane occurs by a factor of $\rho=n_2/n_1 $. As a
consequence the central frequency is reduced. This is of course not
surprising due to the symmetry of Maxwell's equations under the time
reversal operation.

In Fig. \ref{squarelens} the backreflections at the interfaces, when
the pulse enters the structure and when it leaves it, are not shown
for simplicity, but it is clear from the discussions of the previous
sections that for sharp interfaces of the structure they are always
present.

\begin{figure}
\includegraphics{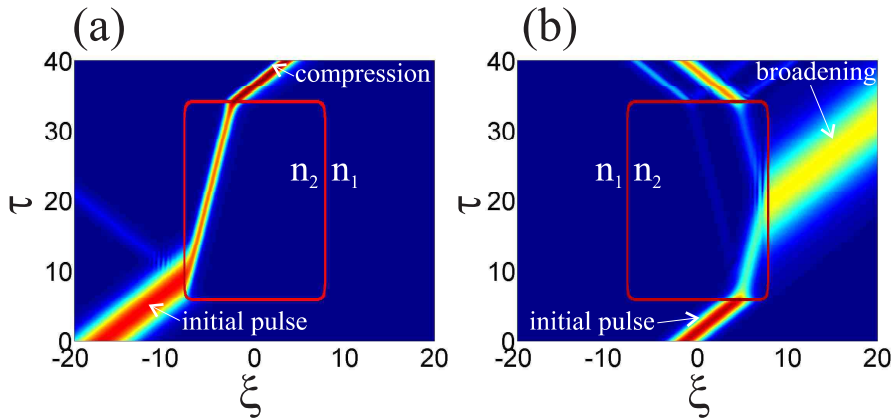}
\caption{\label{squarelenssimul} (Color online) Simulation showing
pulse propagation and scattering by a spatiotemporal lens in the
$(z,ct)$-plane which separates two regions with uniform refractive
indices $n_{1}$ and $n_{2}$, as schematically described by Fig.
\ref{squarelens}. (a) A broad pulse [$w=3$ in Eq.(\ref{pulseform})]
hits the rectangle on the leftmost side, and pulse compression is
achieved. (b) A short pulse [$w=1$ in Eq.(\ref{pulseform})] hits the
structure on the bottom side, and pulse broadening is achieved. For
sake of clarity, only the interface of the rectangle is shown.
Parameters of the simulations are: $n_{1}=1$, $n_{2}=3$, $z_{0}=0$,
$ct_{0}=20\lambda$, $w_{t}=27\lambda$, $w_{z}=15\lambda$,
$m_{t}=m_{z}=20$, and $\lambda$ is an arbitrary scale which we take
equal to the central pulse wavelength. Horizontal axis is
dimensionless space coordinate $\xi=z/\lambda$, while the vertical
axis is dimensionless time coordinate $\tau=ct/\lambda$, and
$\lambda$ is the central pulse wavelength.}
\end{figure}

The scenarios described above are confirmed by direct numerical
simulations, see the panel of Fig. \ref{squarelenssimul}. Fig.
\ref{squarelenssimul}(a) shows the evolution of a spatially and
temporally broad input pulse [$w=3$ in Eq.(\ref{pulseform})]. After
hitting the leftmost side of the rectangular spatiotemporal region,
the pulse emerges from the upper side considerably compressed, both
in spatial and temporal domain. Only the interface of the $n_{2}$
region is indicated for sake of clarity, in order to clearly
distinguish the internal wave propagation.
Fig.~\ref{squarelenssimul}(b) shows the compression dynamics, in
which an input pulse with $w=1$ hits the structure on the bottom
side. The pulse emerging from the right-hand side acquires a
significant broadening. Several waves propagating backwards are
noticeable in Fig.~\ref{squarelenssimul}(a,b), and these are due to
the sharpness of the interface region, which is modeled by
Eq.(\ref{indicelens}) with the specific parameters given in the
caption.

So far we have considered the extreme cases where both interfaces
are space- or time-like. With the formalism developed in the
previous section we can also consider the general case where a pulse
enters a medium with refractive index $n_2$ via an interface
characterized by an angle $\theta_1$ and leaves that medium via
another interface characterized by an angle $\theta_2$. A
straightforward geometrical analysis gives the rescaling factor in
the $(z,ct)$-plane as
\begin{equation}
  \label{eq:compression}
  \rho = \left(\frac{\cos
    \theta_2-n_1\cos\theta_1}{\cos
    \theta_2-n_2\cos\theta_1}\right)\left(\frac{\sin
    \theta_2-n_2\sin\theta_1}{\sin \theta_2-n_1\sin\theta_1}\right)
\end{equation}
which as expected reduces to $\rho=n_1/n_2$ ( $\rho=n_2/n_1$ ) for
the previously considered cases $\theta_1=0$ and $\theta_2=\pi/2$
($\theta_2=0$ and $\theta_1=\pi/2$).

\section{Derivation of a Slow-Variable equation. Comparison with Eq.(\ref{main})}
\label{slow}

\begin{figure}
\includegraphics{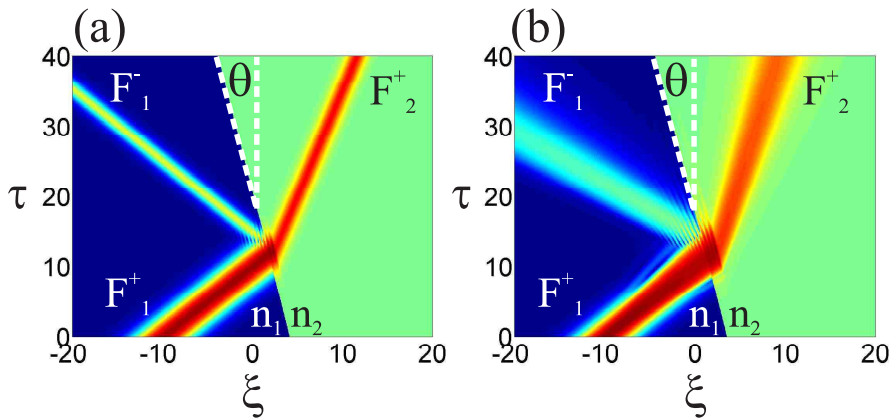}
\caption{\label{diffraction} (Color online) Comparison between pulse
evolution described by (a) the full Eq.(\ref{main}) and (b) the
approximate Eq.(\ref{scalorasveat}), which was derived under
assumption of SVEA in time. The two simulations are carried out
under absolutely identical conditions. Parameters are: $\beta=-0.2$,
$\theta=-0.2027$, $w=1$, $n_{1}=1$ and $n_{2}=3$. In (b) one can
notice an unphysical artificial broadening along the $z$-direction
due to the uncompensated term proportional to the second derivative
in space $\de_{\xi}^{2}$ in Eq.(\ref{scalorasveat}). Horizontal axis
is dimensionless space coordinate $\xi=z/\lambda$, while the
vertical axis is dimensionless time coordinate $\tau=ct/\lambda$,
and $\lambda$ is the central pulse wavelength.}
\end{figure}

From Eq.(\ref{maxwell}) it is possible to derive an equation for the
electric field envelope function $A(z,t)$ only, which is of second
order in the variables $z$ and $t$, by posing
$E(z,t)=[A\exp(ik_{0}z-i\omega_{0}t)+\cc]/2$, where
$k_{0}=\omega_{0}/c$ and $\omega_{0}$ is an arbitrary frequency,
usually and most conveniently chosen to be close to the central
frequency of the input pulse. Using the dimensionless variables
$\tau=ct/\lambda$, $\xi=z/\lambda$, where $\lambda=2\pi
c/\omega_{0}$, and substituting the expression of the electric field
$E$ into Eq.(\ref{maxwell}), we obtain:
$$\left( in^{2}-\frac{n\de_{\tau}n}{\pi}
\right)\de_{\tau}A-\frac{n^{2}}{4\pi}(\de^{2}_{\tau}A)+i\de_{\xi}A+\frac{1}{4\pi}\de_{\xi}^{2}A+$$
\eq{scalora1}{\pi\left[
n^{2}-1+\frac{2in\de_{\tau}n}{\pi}-\frac{(\de_{\tau}n)^{2}}{2\pi^{2}}-\frac{n\de_{\tau}^{2}n}{2\pi^{2}}
\right]A=0.} Although Eq.(\ref{scalora1}) is written for the slow
variable $A$, it is {\em exact}, and it is able to handle arbitrary
pulses, and forward and backward propagation at the same time, due
to the fact that it is completely equivalent to the full wave
equation which can be derived from combining the two Maxwell
equations of Eq.(\ref{maxwell}).

It is reasonable to conjecture that the use of the slowly varying
envelope approximation (SVEA) in time in Eq.(\ref{scalora1}) is
valuable in order to reduce the computational complexity of the
problem. However, as we will show shortly, the use of SVEA in time
but not in space is dangerous in that it leads to wrong results for
pulse propagation, and this is especially important for the
dielectric systems we are investigating in this paper, which show
strong variations in the refractive index both in space and time.
The use of SVEA in space in Eq.(\ref{scalora1}) is not permitted in
particular when simulating sharp interfaces and photonic crystals,
and in general in those systems in which the backward propagating
waves are of the same order of magnitude of the forward propagating
ones. Therefore all second derivatives $\de_{\xi}^{2}$ must be
maintained in Eq.(\ref{scalora1}).

Assuming that the central frequency of the pulse $\omega_{0}$ is
much greater than the variation of the envelope function $A$ in
time, we can perform a multiple scale reduction of
Eq.(\ref{scalora1}), obtaining the following first order equation in
time:
\eq{scalorasveat}{in^{2}\de_{\tau}A+i\de_{\xi}A+\frac{1}{4\pi}\de_{\xi}^{2}A+\pi\left[
n^{2}-1+\frac{2in\de_{\tau}n}{\pi} \right]A=0.}
Eq.(\ref{scalorasveat}) is a generalization to time-dependent
refractive index of the propagation equation used in a series of
works \cite{scalora}, where temporal variations of $n$ were not
taken into account. Eq.(\ref{scalorasveat}) shows that the purely
real 'Schroedinger potential' $n^{2}-1$ used in \cite{scalora} is
modified by a purely imaginary term proportional to the time
derivative of the refractive index, namely $i\de_{\tau}n^{2}/\pi$.

Fig. \ref{diffraction} shows the comparison between the scattering
of a pulse by a sharp interface, when calculated by means of the
full Eqs.(\ref{main}) [Fig. \ref{diffraction}(a)] and by means of
the approximate equation Eq.(\ref{scalorasveat}) [see Fig.
\ref{diffraction}(b)], under absolutely identical conditions.
Although the two equations give very similar results for what
concerns the transmission and reflection coefficients, the most
striking difference between the pulse evolution in the two cases is
the presence of a fictitious and unphysical broadening along the
$z$-direction in Fig. \ref{diffraction}(b). This is due to the fact
that the use of SVEA in time but not SVEA in space in
Eq.(\ref{scalorasveat}) leads to an uncompensated 'diffractive' term
proportional to $\de_{\xi}^{2}A$, which is responsible for the
broadening observed in Fig. \ref{diffraction}(b) for all waves
participating in the scattering.

The above considerations provide a justification of the use of the
full Maxwell equations throughout the present paper, instead of
equations using slow variables.

\section{Discussion and Conclusions}
\label{conclusions}

Before concluding, we would like to spend a few words on the
experimental relevance of the topics discussed in the present paper.

Many theoretical and experimental results exist in the literature,
that discuss the interaction of electromagnetic pulses with a plasma
front \cite{berezhiani,avitzour,wilks,lampe,hashimshony,bessarab}.
This interaction has been typically used in practice to shift the
central frequency of an incident pulse by means of the effect
discussed in Section \ref{spacetimelenses}, see Refs. \cite{wilks},
or to convert static electric fields into electromagnetic pulses by
means of a capacitor array \cite{hashimshony}. These effects may
turn out to be extremely useful in the creation of efficient tunable
laser sources \cite{berezhiani}. For instance, in typical
experiments \cite{avitzour}, when a 'probe' pulse is propagating
along the longitudinal coordinate of a semiconductor waveguide, one
can change the refractive index of the material along all its
spatial extension by means of irradiating the plane of the crystal
with another 'source' pulse that creates a electron-hole plasma
throughout the crystal. If the source pulse has a certain angle of
incidence $\alpha$ with respect to the normal to the plane of the
waveguide, then the ionization front of the plasma will move with
velocity $v=c/\sin(\alpha)$ \cite{berezhiani,bessarab}. Note that
for normal incidence ($\alpha=0$), this front moves instantaneously,
and it corresponds to the time-like interface
($\theta\rightarrow\pi/2$, where the general relation between
$\theta$ and $\alpha$ is $\theta+\alpha=\pi/2$) discussed in Section
\ref{nonstationaryinterface}, which follows from a simultaneous
change of refractive index throughout all the spatial extension of
the crystal. This front does not carry any information, and it is
not associated to any moving object, therefore its propagation is
not limited by the speed of light in vacuum. In all these cases,
none of the actual particles in the excited medium is moving faster
than $c$; instead, the front profile only has an apparent motion,
which can be characterized by an effective velocity given by the
above formula, but does not correspond by any means to any real
superluminal motion of particles or information, which is of course
forbidden by special relativity. In this sense, $n(z,t)$ can be a
totally arbitrary function with no relation whatsoever to moving
bodies. In another kind of experiments \cite{bessarab}, short pulses
of the order of nanoseconds are created by irradiating a metal
surface by X-rays at a certain angle of incidence $\alpha$, which
creates a photoelectric current. Again, this current undergoes an
apparent motion at a velocity $v=c/\sin(\alpha)$ and, by means of
superluminal self-phasing of the excited elementary dipoles
(predicted by Carron and Longmire in 1976 \cite{bessarab}), emits a
short electromagnetic pulse at the same angle $\alpha$. It is also
interesting to point out that the processes described briefly above
can also have considerable implications for the physics of
astrophysical plasmas \cite{hashimshony}.

A remark which is important for the present work is that it is
assumed that the physical experiments well approximate a condition
of quasi-one-dimensional propagation of the incident probe pulse,
therefore implying a well-localized confinement of the transverse
profile by means of a strong waveguiding process which does not
allow diffraction, so that the equations we have derived in this
paper, having the transverse ($x$, $y$) and the longitudinal ($z$)
degrees of freedom decoupled, can be safely utilized.

In conclusion, in this paper the physics of light propagation
scattered by nonstationary interfaces has been investigated
analytically and numerically in a unified approach. Interface
transfer matrix and propagation matrix have been found for the
general case of arbitrary velocities of the interface. These
ingredients have been used to construct more complicated
spatiotemporal dielectric structures. In particular, we have first
investigated a spatiotemporal Bragg reflector, where we have given
the generalized condition for the existence of forbidden bandgaps,
and found that the frequency bandgaps close for a critical value of
$\theta=\theta_{cr1}$, with the successive opening of the wavenumber
bandgap at a second critical angle $\theta=\theta_{cr2}$. Secondly
we have explored light scattering by a spatiotemporal lens device,
and the spectral manipulation of pulses has been demonstrated both
analytically and numerically, the most important aspect of which is
a potentially strong pulse compression/broadening.

Future works include the natural extension of the theory given here
to a dispersive susceptibility.

This research has been supported by Science Foundation Ireland (SFI)
and the Irish Research Council for Science, Engineering and
Technology (IRCSET).


\begin{thebibliography}{99}

\bibitem{inhomogeneous} W. C. Chew, {\em Waves and Fields in Inhomogeneous
Media} (Wiley-IEEE Press, 1999).

\bibitem{pochiyeh} P. Yeh, {\em Optical Waves in Layered Media} (Wiley and Sons, N.Y.,
2005).

\bibitem{jackson} J. D. Jackson, {\em Classical Electrodynamics} (Wiley and Sons, New York, 1975).

\bibitem{landau} L. D. Landau and E. M. Lifshits, {\em Electrodynamics of Continuous Media} (Pergamon Press, Oxford, 1960).

\bibitem{yeh} C. Yeh and K. F. Casey, Phys. Rev. {\bf 144}, 665 (1966);
C. Yeh, Phys. Rev. {\bf 167}, 875 (1968); Phys. Rev. E {\bf 48},
1426 (1993).

\bibitem{saca} J. M. Saca, J. Mod. Opt. {\bf 36}, 1367 (1989).

\bibitem{huang} Y. X. Huang, J. Mod. Opt. {\bf 44}, 623 (1997).

\bibitem{visser} M. Visser, Class. Quant. Grav. {\bf 15}, 1767
(1998);  Phys. Rev. Lett. {\bf 85}, 5252 (2000).

\bibitem{leo} U. Leonhardt and P. Piwnicki, Phys. Rev. Lett. {\bf 82},
2426 (1999);  Phys. Rev. A {\bf 60}, 4301 (1999);  Phys. Rev. Lett.
{\bf 84}, 822 (2000);  Phys. Rev. A {\bf 62}, 055801 (2000);  J.
Mod. Opt. {\bf 48}, 977 (2001); Appl. Phys. B {\bf 72}, 51 (2001).

\bibitem{defelice} F. De Felice, Gen. Rel. and Grav. {\bf 2}, 347
(1971).

\bibitem{ostrovsky} L. A. Ostrovsky, Sov. Phys. Uspekhi {\bf 18}, 452 (1975).

\bibitem{morgenthaler} F. R. Morgenthaler, IRE Trans. Microwave Theory Tech. {\bf MTT-6},
167 (1958).

\bibitem{felsen} L. B. Felsen and G. M. Whitman, IEEE Trans. Antennas Propagat. {\bf
AP-18}, 242 (1970).

\bibitem{fante} R. L. Fante, IEEE Trans. Antennas Propagat. {\bf
AP-19}, 417 (1971).

\bibitem{nonstationary} A. B. Shvartsburg, Physics Uspekhi {\bf 48},
797 (2005).

\bibitem{nakahara} M. Nakahara, {\em Geometry, Topology, and Physics} (2nd ed., Taylor and
Francis, London, 2003).

\bibitem{scalora} J. P. Dowling, M. Scalora, M. J. Bloemer and
C. M. Bowden, J. Appl. Phys {\bf 75}, 1896 (1994); M. Scalora, J. P.
Dowling, C. M. Bowden and M. J. Bloemer, Phys. Rev. Lett. {\bf 73},
1368 (1994); M. Scalora and M. E. Crenshaw, Opt. Commun. {\bf 108},
191 (1994); M. Scalora, J. P. Dowling, C. M. Bowden and M. J.
Bloemer, J. Appl. Phys. {\bf 76}, 2023 (1994); M. Scalora, M. J.
Bloemer, A. S. Manka, J. P. Dowling, C. M. Bowden, R. Viswanathan
and J. W. Haus, Phys. Rev. A {\bf 56}, 3166 (1997).

\bibitem{kronigpenney} S. Mishra and S. Satpathy, Phys. Rev. B {\bf
68}, 045121 (2003).

\bibitem{berezhiani} V. I. Berezhiani, S. M. Mahajan, and R.
Miklaszewski, Phys. Rev. A {\bf 59}, 859 (1999).

\bibitem{avitzour} Y. Avitzour, I. Geltner and S. Suckewer, J. Phys.
B: At. Mol. Opt. Phys. {\bf 38}, 779 (2005).

\bibitem{wilks} S. C. Wilks, J. M. Dawson and W. B. Mori, Phys. Rev. Lett. {\bf
61}, 337 (1988); R. L. Savage, C. Joshi and W. B. Mori, Phys. Rev.
Lett. {\bf 68}, 946 (1992); I. Geltner, Y. Avitzour and S. Suckewer,
Appl. Phys. Lett. {\bf 81}, 226 (2002).

\bibitem{lampe} M. Lampe, E. Ott and J. H. Walker, Phys. Fluids {\bf
21}, 42 (1978).

\bibitem{hashimshony} D. Hashimshony, A. Zigler and K. Papadopoulos,
Phys. Rev. Lett. {\bf 86}, 2806 (2001); C. H. Lai, R. Liou, T. C.
Katsouleas, P. Muggli, R. Brogle, C. Joshi and W. B. Mori, Phys.
Rev. Lett. {\bf 77}, 4764 (1996).

\bibitem{bessarab} N. J. Carron and C. L. Longmire, IEEE Trans. Nucl. Sci. {\bf NS-23}, 1897 (1976);
A. V. Bessarab, A. A. Gorbunov, S. P. Martynenko and N. A. Prudkoy,
IEEE Trans. Plasma Sci. {\bf 32}, 1400 (2004).


\end{thebibliography}
\end{document}